\documentclass[10pt,a4paper]{article}
\usepackage[utf8]{inputenc}
\usepackage[english]{babel}

\usepackage{amsmath}
\usepackage{amsfonts}
\usepackage{amssymb}

\usepackage{graphicx}

\usepackage{subcaption}

\usepackage{wrapfig}
\usepackage{pb-diagram}

\usepackage[colorlinks,citecolor=blue,urlcolor=blue,linkcolor=blue]{hyperref}

\usepackage[left=2cm,right=2cm,top=2cm,bottom=2cm]{geometry}

\usepackage{authblk}

\title{Inflation in simple one-loop effective potentials of perturbative quantum gravity}
\author[1,2]{\href{https://orcid.org/0000-0001-9326-6905}{A. Arbuzov} \thanks{\href{mailto:arbuzov@theor.jinr.ru}{arbuzov@theor.jinr.ru}} }
\author[2]{\href{https://orcid.org/0009-0002-3648-875X}{D. Kuznetsov}
\thanks{\href{mailto:kuzda.18@uni-dubna.ru}{kuzda.18@uni-dubna.ru}} }
\author[3]{\href{https://orcid.org/0000-0001-7099-0861}{B. Latosh} \thanks{\href{mailto:latosh.boris@ibs.re.kr}{latosh.boris@ibs.re.kr}} }
\author[4]{\href{https://orcid.org/0009-0009-6814-0986}{V. Shmidt} \thanks{\href{mailto:shmidt.v@phystech.edu}{shmidt.v@phystech.edu}} }
\affil[1]{Bogoliubov Laboratory of Theoretical Physics, JINR, Dubna, 141980, Russia}
\affil[2]{Dubna State University, Universitetskaya str. 19, Dubna, 141982, Russia}
\affil[3]{Particle Theory  and Cosmology Group, Center for Theoretical Physics of the Universe, Institute for Basic Science (IBS), Daejeon, 34126, Korea}
\affil[4]{Moscow Institute of Physics and Technology, Dolgoprudny, 141701, Russia}
\date{CTPU-PTC-23-30}

\begin{document}

\maketitle

\abstract{
  We study inflation in scalar-tensor perturbative quantum gravity driven by a one-loop effective potential. We consider effective potentials generated by three models. The first model describes a single scalar field with a non-vanishing mass. The second model describes a massless scalar field with non-minimal coupling to the Einstein tensor. The third model generalises the Coleman-Weinberg model for the gravitational case. The first model can be consistent with the observational data for $N\sim 70$ e-foldings. The second model can be consistent with the observational data for $N \sim 40$ e-foldings. We did not find parameters that make the generalised Coleman-Weinberg model consistent with the observational data. We discuss the implications of these results and ways to improve them with other terms of effective action.
}

\section{Introduction}

The inflationary phase of the universe's expansion is essential to cosmological evolution. It resolves several serious problems that otherwise arise in cosmological models \cite{Starobinsky:1980te,Linde:1983gd,Gorbunov:2011zzc,Rubakov:2017xzr,Mukhanov:2005sc}. Contemporary empirical data on the cosmic microwave background radiation and the large-scale structure of the universe provide some constraints on the inflationary parameters \cite{Planck:2018jri,WMAP:2012nax,BICEP:2021xfz,Paoletti:2022anb,Planck:2013jfk}. However, they are still insufficient to select a single model that best fits the full range of data. As a result, there is a large number of inflationary models constructed on very different foundations \cite{Martin:2013tda,Linde:2014nna,Nojiri:2017ncd,Kobayashi:2011nu,Bezrukov:2010jz,Barvinsky:2008ia,Oikonomou:2021kql,Oikonomou:2022bqb,Oikonomou:2024tjf}, see also \cite{Odintsov:2023weg} for a recent review. 

One of the most promising approaches is considering inflation as an effect caused by quantum corrections. Several models implement this approach directly or indirectly. Perhaps the best known is the Starobinsky inflation \cite{Starobinsky:1980te}. This model modifies general relativity by including a quadratic curvature term that drives inflation. This model can be mapped onto a scalar-tensor model with a non-trivial scalar field potential. In this parameterisation, inflation is driven by the well-known slow-roll mechanism \cite{Linde:1983gd,Liddle:1994dx}. The squared curvature term is strongly related to quantum gravitational corrections. At the one-loop level, quantum gravity universally generates operators that are quadratic in curvature \cite{tHooft:1974toh,Deser:1974cz,Goroff:1985sz}. On the cosmological background, these operators are reduced to full derivatives, except for the curvature-squared operator, which is present in the Starobinsky model. This example further motivates us to explore inflation's roots in pure quantum effects.

Effective action formalism provides a tool for constructing inflationary models from quantum field theory considerations. The formalism itself, as we will discuss below, constructs an effective action $\Gamma$ which depends on the vacuum expectation values of a scalar (or any other) field so that it describes an emergent classical dynamics generated by a quantum system \cite{Buchbinder:1992rb,Coleman:1973jx,Jackiw:1974cv,Barvinsky:1985an,Vilkovisky:1984st}. Effective action can be calculated perturbatively by a resummation of one-particle irreducible diagrams. This paper only considers the leading contribution to effective action at the one-loop level, so that we will study one-loop effective actions.

An effective action contains both operators with derivatives and operators without derivatives. The part of the action that does not contain derivatives is usually called the effective potential. As mentioned above, inflation can be driven by a scalar field potential. Consequently, effective one-loop scalar field potentials provide models capable of driving inflation. We should note that inflation can also be driven by a non-minimal scalar field coupling to gravity, which is also present in one-loop effective actions. Discussion of these cases is beyond the scope of this paper.

The main goal of this paper is to investigate simple one-loop effective potentials obtained in perturbative quantum gravity and to determine whether they alone can provide inflationary models consistent with the observational data. We consider models studied in papers \cite{Arbuzov:2020pgp,Arbuzov:2021yai}. These models describe effective potentials generated by the simplest scalar-tensor models. We show that all discussed models can drive inflation but are inconsistent with the observational data for the simplest case of $N=60$ e-foldings. The first model favours a large number of e-foldings $N\sim 70$, while the second model favours a smaller number $N \sim 40$. The third model is strongly inconsistent with the observational data, and an agreement cannot be achieved for a wide range of model parameters.

The paper is organised as follows. In Section \ref{Section_Models}, we discuss the models under study in more detail. We cover the basics of perturbative quantum gravity and effective action. We then discuss the effective potentials calculated in \cite{Arbuzov:2020pgp,Arbuzov:2021yai}. In Section \ref{Section_Inflation}, we discuss inflation parameters that can be calculated in models with such effective potentials and present the results of such calculations. In Section \ref{Section_Conclusion}, we discuss the physical implications of these results and the further development of this approach. 

\section{One loop effective potentials}\label{Section_Models}

We consider models based on perturbative quantum gravity. Here we briefly review the formalism; a more detailed discussion presents in \cite{Latosh:2022ydd,Latosh:2023zsi,Latosh:2020ysu,Burgess:2003jk}. The central premise of quantum gravity is an association of weak quantum gravitational effects with small metric perturbations propagating in flat spacetime. The following perturbative expansion gives the total metric:
\begin{align}
  g_{\mu\nu} = \eta_{\mu\nu} + \kappa\, h_{\mu\nu} \, .
\end{align}
Here $\kappa$ is the gravitational coupling related to Newton's constant $\kappa^2 = 32 \,\pi\, G$. This is a finite expansion, but it produces an infinite series in $\kappa$ for the Einstein-Hilbert action. The path integral formalism gives a way to compute matrix elements with the generating functional:
\begin{align}
  \begin{split}
    \mathcal{Z}[J] & = \int \mathcal{D}[g] \, \exp\left[ i\, \mathcal{A}[g]~\right] = \int \mathcal{D}[\eta + \kappa\, h] \,\exp\left[ i\,\mathcal{A}[\eta + \kappa \, h] + i \, J \cdot h \right] \\
    & = \int \mathcal{D}[h] \,\exp\left[ i\, h_{\mu\nu}\mathcal{D}^{\mu\nu\alpha\beta}\square \, h_{\alpha\beta} + i \, J \cdot h + \mathcal{O}\left(\kappa\right) \right].
  \end{split}
\end{align}
Here $\mathcal{A}$ is the microscopic action defining the structure of the theory, $J$ are formal external currents, and we have omitted higher order terms describing the graviton interaction. This functional generates matrix elements by taking an appropriate derivative:
\begin{align}
  \langle 0 \lvert h_{\mu\nu}(x) \cdots h_{\alpha\beta}(y) \rvert 0 \rangle = \cfrac{1}{\mathcal{Z}[J]} \cfrac{\delta}{\delta\,i\,J^{\mu\nu}(x)} \cdots \cfrac{\delta}{\delta\,i\,J^{\alpha\beta}(y)} \,\mathcal{Z}[J] \Bigg|_{J=0}. 
\end{align}

The effective action is constructed as follows. One starts with the connected generating functional $W$:
\begin{align}
  \mathcal{Z}[J] = \exp[ i\, W[J]].
\end{align}
Derivatives of the generating functional $\mathcal{Z}$ generate matrix elements corresponding to all possible Feynman graphs, but the connected generating functional $W$ only generates matrix elements corresponding to connected graphs. The effective action $\Gamma$ is a Legendre transformation of $W$:
\begin{align}
  \Gamma[\varphi] = W[J] - \varphi\,\cdot J.
\end{align}
This quantity is called effective action for two reasons. First, the derivative of the effective action by $J$ shows that $\varphi$ is the classical value of the quantum scalar field $\phi$:
\begin{align}
  \varphi = \cfrac{\delta}{\delta J} \, W[J] = \cfrac{1}{\mathcal{Z}[J]} \, \cfrac{\delta}{\delta\,i\,J} \, \mathcal{Z}[J] = \langle 0 \lvert \phi \rvert 0 \rangle .
\end{align}
Second, the derivative of the effective action by $\varphi$ vanishes when there are external currents $J$:
\begin{align}
  \cfrac{\delta \Gamma}{\delta \varphi} = - J \overset{J \to 0}{\to} 0.
\end{align}
These properties match the properties of a classical action describing a classical system, which allows one to consider the effective action as a proper action describing the classical dynamics of a given quantum system.

This paper considers the effective potentials calculated in the papers \cite{Arbuzov:2020pgp,Arbuzov:2021yai}. These models provide minimal modifications of general relativity, which makes them the simplest natural candidates. As mentioned above, we will discuss only the effective potential of the scalar field rather than the complete effective action for the following reasons. First, inflation can be driven either by a scalar field potential or by the non-minimal coupling of the scalar field to gravity. The one-loop effective action for most scalar-tensor models includes both the effective potential and the non-minimal coupling. It is essential to study their influence separately to distinguish their roles. Second, in models that allow both a scalar field potential and a non-minimal coupling, two independent inflationary phases can occur \cite{Avdeev:2021von,Avdeev:2022ilo}. The non-minimal coupling can have a non-trivial influence on the inflationary phase driven by the scalar field potential. Consequently, it is crucial to evaluate inflationary parameters without non-minimal couplings to see if they can improve the consistency with the empirical data. For this reason, we omit any discussion of non-minimal couplings generated at the one-loop level and focus only on the one-loop effective potential.

The following microscopic action gives the first model:
\begin{align}
  \mathcal{A}_\text{I} = \int d^4 x \sqrt{-g} \left[ -\cfrac{2}{\kappa^2} \,R - \cfrac{1}{2}\,\phi \left(\square + m^2\right) \phi \right].
\end{align}
It is the simplest model describing a single massive scalar field minimally coupled to gravity. Without gravity, the model cannot develop an effective potential within standard quantum field theory because the scalar field lacks the interaction sector. Within perturbative quantum gravity, it develops the following effective potential at the one-loop level:
\begin{align}\label{V_I}
  \begin{split}
    V_\text{I} =& -m^4 \frac{\ln{2}}{64 \pi^2} + \frac{m^2}{2} \left[1+\frac{m^2 \kappa^2}{32 \pi^2}(1+\ln{4}) \right] \varphi^2 \\
    & +\cfrac{m^4}{128 \pi^2} \left( 1 - 2\, \kappa^2 \varphi^2 -\sqrt{ 1 - 4\,\kappa^2 \varphi^2} \right) \ln{\left[1-\sqrt{1 - 4\, \kappa^2 \varphi^2} \right]} \\
    & +\cfrac{m^4}{128 \pi^2}\left( 1 - 2\, \kappa^2 \varphi^2 + \sqrt{ 1 - 4\,\kappa^2 \varphi^2} \right) \ln{\left[1+\sqrt{1 - 4\, \kappa^2 \varphi^2} \right]} .
  \end{split}
\end{align}
Before renormalisation, the potential had a single ultraviolet divergence in the mass term. It is renormalised at $\varphi=0$ so that the classical field $\varphi$ has the same mass as the quantum field $\phi$. The potential has a single local minimum at $\varphi=0$ and local maxima deep in the Planck region. The equation defining the positions of these maxima is transcendental, so we could not find an explicit formula defining their position. The potential increases monotonically from $\varphi=0$ to the local maxima, after which it decreases indefinitely. The position of the maxima is far in the Planck region, so they naturally mark the most general applicability of the model. 

The following action gives the second model
\begin{align}
  \mathcal{A}_\text{II} = \int d^4 x\sqrt{-g} \left[ - \cfrac{2}{\kappa^2}\, R - \cfrac12\, \phi\, \square\,\phi - \cfrac{\lambda}{2}\,R \, \phi^2 \right].
\end{align}
Here $\lambda$ is a dimensionless coupling. The new non-minimal coupling is the only one that describes the three-particle interaction, contributes to the effective potential and does not introduce higher derivative terms in the field equations. We exclude the scalar field mass term to separate its contribution from this non-minimal coupling. The model produces the following effective potential after renormalisation:
\begin{align}\label{V_II}
  V_\text{II} = \cfrac{m^2}{3\,\kappa^2\,\lambda^2}\, \ln\left[1 + \cfrac32\,\lambda^2\,\kappa^2\,\varphi^2 \right].
\end{align}
This potential has a single minimum at $\varphi=0$ and grows monotonically with $\varphi$.

Unlike the previous case, the potential has infinite UV-divergent terms parameterised with a single coupling. This provides the basis for a consistent renormalisation of this model. Although we have to introduce an infinite number of counterterms, all of them are defined by a single constant. The potential is renormalised at $\varphi=0$, where its mass is set to a finite value $m$. It should also be noted that the UV divergence discussed was a power-law divergence regularised in a cut-off scheme. Consequently, the potential is expected to be sensitive to the UV structure of the theory. In other words, the mass of the classical field is highly sensitive to possible UV extensions of the theory.

Finally, the third model is a generalisation of the well-known Coleman-Weinberg model \cite{Coleman:1973jx}:
\begin{align}
  \mathcal{A}_\text{III} = \int d^4 x \sqrt{-g} \left[ -\cfrac{2}{\kappa^2}\,R - \cfrac12\,\phi (\square + m^2) \phi - \cfrac{\lambda}{4}\,\phi^4\right].
\end{align}
The model is important because it contains a simple scalar field self-interaction, contributing to the effective potential. Without gravity, the model has good renormalisation behaviour. Because of this, the corresponding effective potential will contain two competing contributions, and it would be possible to determine their influence on inflationary scenarios. The corresponding leading order one-loop corrections after renormalisation are given by the following expression:
\begin{align}\label{V_III}
  \begin{split}
    V_\text{III} =& \ln\left[ 1 + \cfrac{\frac{\lambda}{2}\,\varphi^2}{m^2} \right]\left\{ \cfrac{m^4}{64\pi^2} + \cfrac{m^2}{2}\,\varphi^2\,\cfrac{\lambda-2\,m^2\kappa^2}{32\pi^2} + \cfrac{\lambda}{4!}\,\varphi^4\,\cfrac{3\,\lambda -8\,m^2\kappa^2}{32\pi^2} - \cfrac{\kappa^2}{6!}\,\varphi^6\,\cfrac{5\,\lambda^2}{8\pi^2} \right\}\\
    & + \cfrac{m^2}{2}\left(1-\cfrac{\lambda}{64\pi^2}\right)\,\varphi^2 + \cfrac{\lambda}{4!}\left( 1 - \cfrac{3\,(3\lambda - 8\,\kappa^2m^2)}{64\pi^2}\right)\varphi^4 \\
    & + \cfrac{g}{6!} \left( 1 - \cfrac{15\,\lambda^2}{32\pi^2}\,\cfrac{\lambda - 2\, m^2\kappa^2}{g\,m^2}\right) \,\varphi^6 .
  \end{split}
\end{align}
This case is similar to the first model. First, the potential has a single local minimum at $\varphi=0$, and it grows with $\varphi$ until it reaches its maximum and then approaches $-\infty$. Second, the potential has ultraviolet divergences in the mass and $\varphi^4$ interaction terms. These divergences can be subtracted by the standard scheme and renormalised to the corresponding values of the microscopic action. In contrast to the first case, the potential also has a divergence in the $\varphi^6$ interaction term, which is missing in the microscopic action. Nevertheless, we subtract the corresponding divergence and normalise the potential to a new unknown coupling $g$ with mass dimension $-2$.

In summary, the three models discussed provide minimal modifications of general relativity with a scalar field. The first model introduces a single scalar field minimally coupled to gravity, the second introduces a scalar field with non-minimal coupling, and the last introduces a minimally coupled scalar field with self-interaction. The simplicity of these models gives them a reasonable degree of universality because one can expect similar behaviour in various other models. For example, the contribution to the effective potential generated in the first model will be present in any model containing a massive scalar field. The corresponding inflation scenarios, in turn, provide a tool for making reasonable assumptions about inflation parameters in a broad class of models.

\section{Inflationary parameters}\label{Section_Inflation}

This paper only considers inflation within scalar-tensor gravity models with minimal coupling and scalar field potential. A general action of such a theory is given by the following:
\begin{align}
  S = \int d^4 x \sqrt{-g} \, \left[ -\cfrac{2}{\kappa^2} \, R + \cfrac12\, g^{\mu\nu} \,\nabla_\mu \varphi \,\nabla_\nu \varphi - V(\varphi)\right].
\end{align}
For simplicity, we will only consider a spatially flat Friedmann-Robertson-Walker universe described by the following metric:
\begin{align}
  ds^2 = dt^2 - a^2(t) \left[ dx^2 + dy^2 + dz^2 \right].
\end{align}
We consider only uniform scalar fields that depend only on time coordinate $\varphi=\varphi(t)$. This gives us the familiar field equations:
\begin{align}
    \begin{split}
    & \cfrac{12}{\kappa^2}\, H^2 = \frac{1}{2} \, \dot{\varphi}^2+V(\varphi), \\
    & \ddot{\varphi} + 3 \, H \, \dot{\varphi} + V'(\varphi)=0 .
    \end{split}
\end{align}

The model enters the inflationary expansion when the Hubble parameter $H=\dot{a}/a$ is constant, which is equivalent to the following conditions on the Hubble parameter \cite{Liddle:1994dx}:
\begin{align}
  \cfrac{\dot{H}(t)}{H^2(t)} & \ll 1 \,, & \cfrac{\ddot{H}(t)}{H^3(t)} & \ll 1.
\end{align}
In the simple inflationary model discussed, they are reduced to the slow roll conditions for the potential:
\begin{align}
  \left( \cfrac{V'(\varphi)}{V(\varphi)}\right)^2  & \ll 1\,, & \cfrac{V''(\varphi)}{V(\varphi)} & \ll 1.
\end{align}
In turn, one can define the so-called slow roll parameters
\begin{align}
  \varepsilon &=\frac{2}{3\, \kappa^2} \,\left( \cfrac{V'(\varphi)}{V(\varphi)} \right)^2 \,, & \eta & = \frac{4}{3 \, \kappa^2} \left\lvert \cfrac{V''(\varphi)}{V(\varphi)} \right\rvert.
\end{align}

The model enters the slow roll regime when both $\eta$ and $\varepsilon$ are small and exists in the slow roll regime then one of these parameters becomes larger than $1$. Consequently, the time $t_\text{end}$ at which inflation ends are defined by the corresponding condition:
\begin{align}
  \eta(t_\text{end})  =1\quad \text{ or } \quad  \varepsilon(t_\text{end}) =1.
\end{align}
The inflation must last for at least $50$ e-foldings to be consistent with the observational data. In the slow roll approximation, the number of e-foldings is given by the following analytical expression:
\begin{align}
  N (t_\text{end}, t_\text{init}) =-\cfrac{\kappa^2}{4} \, \int\limits_{\varphi_{\rm{init}}}^{\varphi_{\rm{end}}} \frac{V(\varphi)}{ V'(\varphi)} \,d \varphi , \quad \varphi_{\rm{init}}=\varphi(t_{\rm{init}}),\quad  \varphi_{\rm{end}}=\varphi(t_{\rm{end}}) .
\end{align}
We use this condition to determine the time $t_\text{init}$ at which inflation began. We mainly focus on the case of $N=60$ e-foldings but also consider some deviations to study if the discussed models can be consistent with the observational data.

We use the tensor-to-scalar ratio $r$ and the scalar tilt of the scalar perturbation spectrum $n_s$ to perform the verification by the empirical data. In the class of models discussed, these parameters are given by the explicit formulae \cite{Granda:2019wpe,Granda:2019jqy}:
\begin{align}
  r = 48 \, \varepsilon, \quad n_s- 1 = 6\, ( \eta - 3\, \varepsilon ).
\end{align}
These parameters are well constrained by the present observational data \cite{Planck:2018jri,WMAP:2012nax,BICEP:2021xfz,Paoletti:2022anb,Planck:2013jfk}:
\begin{align}
r< 0.032, \quad n_s=0.9663 \pm 0.0041 .
\end{align}
This allows us to check whether these models are consistent with the observational data.

\subsection{First model}

Let us start with the first model with the effective potential \eqref{V_I}. The following sophisticated expressions give the slow roll parameters:
\begin{align}
    \begin{split}
    \varepsilon =& \cfrac{32 \, \varphi^2}{3\,\kappa^2\, (1- 4\, \kappa^2 \varphi^2)} \Bigg[ \Bigg\{ \kappa^2 m^2 \left( 1 + \sqrt{ 1 - 4 \, \kappa^2 \varphi^2 } \right) \ln \left( 1 + \sqrt{ 1 - 4 \, \kappa^2 \varphi^2} \right) \\
    & - \kappa^2\, m^2 \, \left( 1 - \sqrt{ 1 - 4\, \kappa^2 \varphi^2} \right) \ln \left( 1 - \sqrt{ 1 - 4 \, \kappa^2 \varphi^2} \right) - \sqrt{ 1 - 4\, \kappa^2 \varphi^2} \left(\kappa^2 m^2 \ln 4 + 32 \pi ^2\right) \Bigg\} \Bigg/ \\
    & \Bigg\{ 64 \, \pi^2 \phi ^2 + m^2 \big( \kappa^2 \varphi^2 ( 2 + \ln 16 )-\ln 4 \big)  + m^2 \left( 1 - 2\, \kappa^2 \varphi^2 + \sqrt{ 1 - 4 \, \kappa^2 \varphi^2} \right) \ln \left( 1 + \sqrt{ 1 - 4\, \kappa^2 \varphi^2} \right)  \\
    & \hspace{10pt} + m^2 \left( 1 - 2 \, \kappa^2 \varphi^2 - \sqrt{1 - 4\, \kappa^2 \varphi^2} \right) \ln \left( 1 - \sqrt{1 - 4\, \kappa^2 \varphi^2 } \right) \Bigg\} \Bigg]^2 ,
    \end{split}
\end{align}
\begin{align}
    \begin{split}
        \eta  = & \cfrac{16}{3\,\kappa^2\, (1 - 4\, \kappa^2 \varphi^2)^{3/2}} \Bigg[ \Bigg\{ \kappa^2 m^2 \left( 1 + ( 1 - 4 \, \kappa^2 \varphi^2)^{3/2} \right) \ln \left( 1 - \sqrt{ 1 - 4 \, \kappa^2 \varphi^2}\right) \\
        & - \kappa^2 m^2 \left( 1 - ( 1 - 4 \, \kappa^2 \varphi^2)^{3/2} \right) \ln \left( 1 - \sqrt{ 1 - 4 \, \kappa^2 \varphi^2} \right) \\
        & + \sqrt{ 1 - 4 \, \kappa^2 \varphi^2} \Big( - 32 \, \pi^2 (1 - 4\, \kappa^2 \varphi^2 ) + \kappa^2 m^2 \left( 4\, \kappa^2 \varphi^2 ( \ln 4 - 2 ) - \ln 4 \right) \Big) \Bigg\}\Bigg/\\
        & \Bigg\{ m^2 \Big( \ln 4 - 2 \, \kappa^2 \varphi^2 ( 1 + \ln 4 ) \Big) - m^2 \left( 1 - 2 \, \kappa^2 \varphi^2 + \sqrt{ 1 - 4 \, \kappa^2 \varphi^2} \right) \ln \left( 1 + \sqrt{ 1 - 4 \, \kappa^2 \varphi^2} \right) \\
        & \hspace{10pt} - m^2 \left( 1 - 2\, \kappa^2 \varphi^2 - \sqrt{ 1 - 4\, \kappa^2 \varphi^2} \right) \ln \left( 1 - \sqrt{ 1 - 4 \,\kappa^2 \varphi^2} \right) - 64\, \pi^2 \varphi^2 \Bigg\} \Bigg] .
    \end{split}
\end{align}
\noindent Inflation can occur only when both of these parameters are small. Because of the sophisticated form of the potential and slow roll parameters, one can hardly draw meaningful conclusions about inflation based on these expressions alone. We analyse the parameter space of the model to draw comprehensive conclusions.

\begin{figure}
  \centering

  \begin{minipage}[b]{0.45\textwidth}
    \centering
    \begin{subfigure}[b]{\textwidth}
      \includegraphics[width=\textwidth]{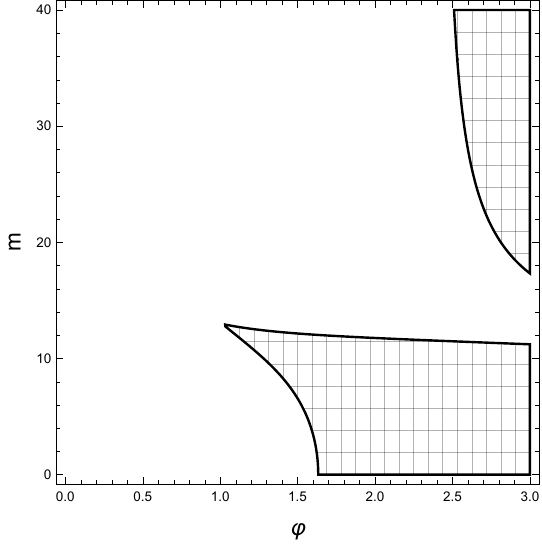}
    \end{subfigure}
  \end{minipage}
  \hfill
  \begin{minipage}[b]{0.45\textwidth}
    \centering
    \begin{subfigure}[b]{\textwidth}
      \includegraphics[width=\textwidth]{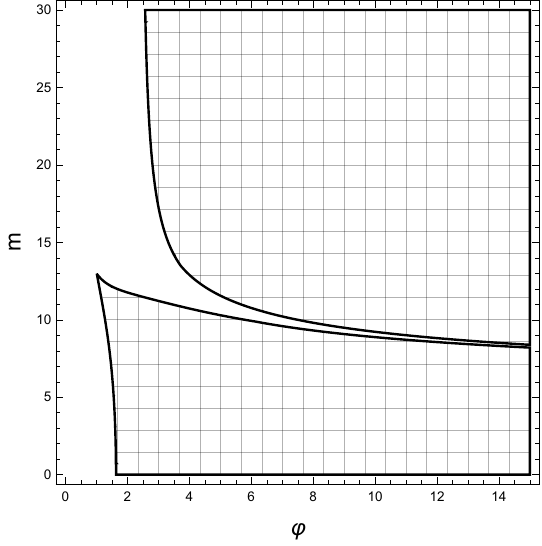}
    \end{subfigure}
  \end{minipage}

  \caption{Potential $V_I$. Parameter space of the model. The slow roll parameters are small in the shaded region. Parameters $\varphi$ and $m$ are given in $\kappa$ units.} \label{fig:VIParameters}
\end{figure}

We present plots showing the region of the model's $(\varphi,m)$ parameter space where the slow roll parameters are small in Figure \ref{fig:VIParameters}. In these plots dimensional parameters $\varphi$ and $m$ are given in the $\kappa$ units, namely, $\kappa\,\varphi$ and $\kappa\,m$ are the corresponding values. In the following analysis, we will also use dimensional variables for the sake of simplicity:
\begin{align}
  \tilde\varphi &= \kappa \, \varphi \,, & \tilde{m} = \kappa\, m \,.
\end{align}

The suitable region of the parameter space has a narrow branch with small slow roll parameters. Nevertheless, a realistic inflationary scenario can hardly occur because the branch is too narrow to embed $60$ e-foldings. Moreover, the branch takes the region with small values of field $\tilde\varphi$ and large values of masses $\tilde{m} > 10$, which makes the setup marginal. Because of this, we will not study scenarios taking place there.

The presence of $\ln$-functions in the potential \eqref{V_I} makes obtaining analytic expressions for most inflationary scenarios impossible. We will use numerical calculations to study inflation in most cases. However, there are two limiting cases where analytic expressions can be obtained. These are the massless limit $m\to 0$ and the heavy mass limit $m\to\infty$. In both of these limits, the effective potential losses its physical meaning: in the massless limit it vanishes, and in the heavy mass limit it becomes infinite. On the contrary, expressions for the slow roll parameters remain finite and well-defined. This does not imply neither that inflation can occur without a potential nor that it can occur with an infinite potential. It only shows that the inflationary parameters show a universal behaviour with a weak mass dependence in these limiting cases.

We begin with the massless limit. In that limit, slow roll parameters read:
\begin{align}
  \varepsilon = \eta = \cfrac{8}{3\,\tilde\varphi^2} \,.
\end{align}
In turn, the expression defining the number of e-foldings is
\begin{align}
  N (t_\text{end}, t_\text{init}) = - \cfrac{1}{8}\, \int\limits_{\tilde\varphi_{\rm{init}}}^{\tilde\varphi_{\rm{end}}}\, \tilde\varphi \,d \tilde\varphi .
\end{align}
These expressions match the case of a model with $\varphi^2$ potential, so we obtain the well-known relations between the inflationary parameters $r$, $n_s$ and the number of e-foldings $N$:
\begin{align}
  r & = \cfrac{48}{6\,N + 1} \, , & n_s & = \cfrac{6\,N -11}{6\,N+1} \,.
\end{align}
Consequently, this limiting case cannot fit the data because it develops $n_s$ consistent with observation for $53 \lesssim N \lesssim 67 $, but $r$ is consistent for $ N \gtrsim 250$. 

\begin{figure}[h]
    \centering
    \begin{subfigure}{0.45\textwidth}
        \centering
        \includegraphics[width=\textwidth]{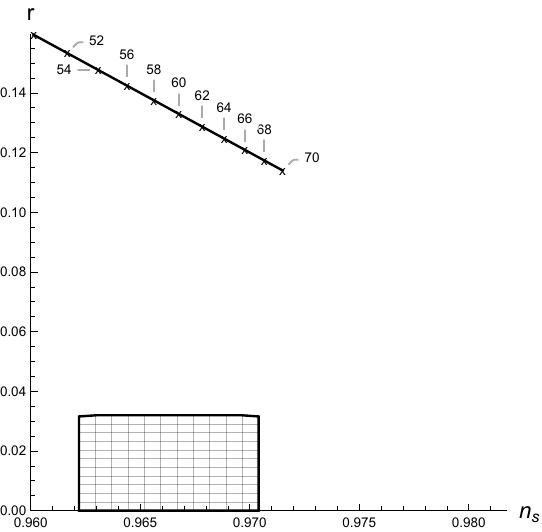}
        \caption{Massless limit.}
        \label{fig:VILimit1}
    \end{subfigure}
    \hfill
    \begin{subfigure}{0.45\textwidth}
        \centering
        \includegraphics[width=\textwidth]{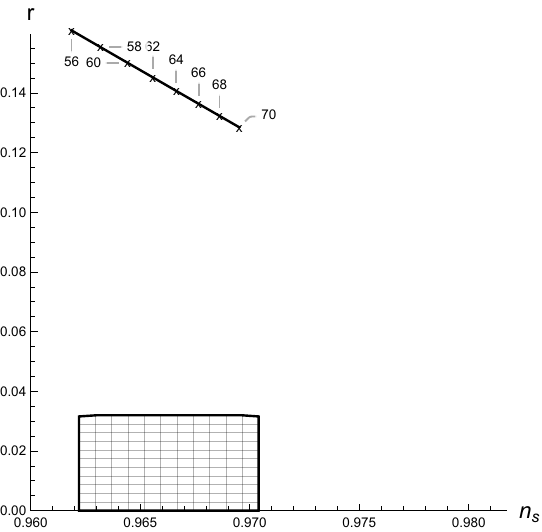}
        \caption{Heavy mass limit.}
        \label{fig:VILimit2}
    \end{subfigure}
    \caption{Potential $V_I$. Inflationary parameters for the limiting cases. Plot markers correspond to different numbers of e-folding. The shaded region marks parameters consistent with the observational data.}
    \label{fig:VILimits}
\end{figure}

For any given $N$, they cannot simultaneously get both $r$ and $n_s$ to be consistent with the observational data. The scalar tensor tilt is consistent with observations for $52.7434 < N < 67.4009$, while the tensor-to-scalar ratio is consistent for $N>249.833$. Figure \ref{fig:VILimit1} presents the plot showing the model's behaviour.

We proceed with the heavy mass limit. Obtaining explicit analytic expressions for the slow roll parameters for this case is possible. However, they are sophisticated and can hardly be analysed directly, so we omit them for briefness. They also contain multiple $\ln$ functions, so obtaining analytic expressions for boundary field values and the number of e-folding is impossible. We study the model with numerical methods and found that it experiences a similar behaviour. Namely, for $56 \lesssim N \lesssim 70$ e-foldings, the model generates $n_s$ consistent with the observational data, but the tensor-to-scalar ratio remains large. The corresponding plot is given in Figure \ref{fig:VILimit2}.

\begin{figure}
  \centering
  
  \begin{minipage}[b]{0.49\textwidth}
    \centering
    \begin{subfigure}[b]{\textwidth}
      \includegraphics[width=\textwidth]{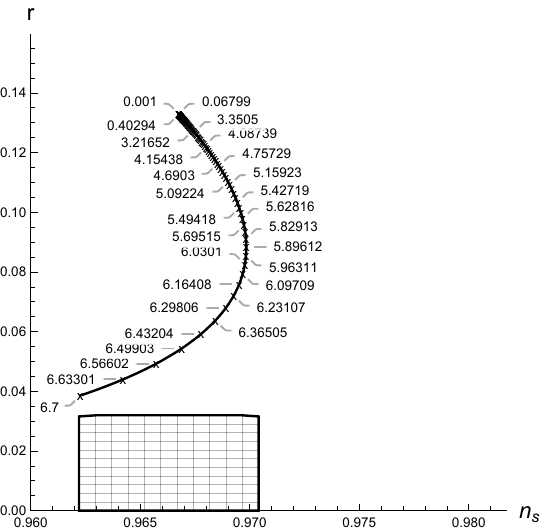}
      \caption{$N=60$}
    \end{subfigure}

    \begin{subfigure}[b]{\textwidth}
      \includegraphics[width=\textwidth]{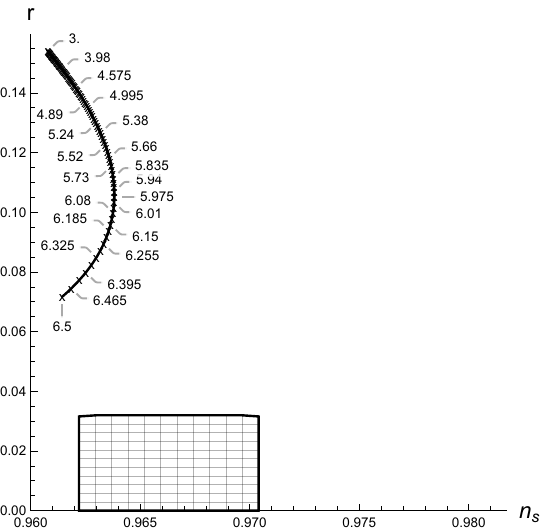}
      \caption{$N=50$}
    \end{subfigure}
  \end{minipage}
  \hfill
  \begin{minipage}[b]{0.49\textwidth}
    \centering
    \begin{subfigure}[b]{\textwidth}
      \includegraphics[width=\textwidth]{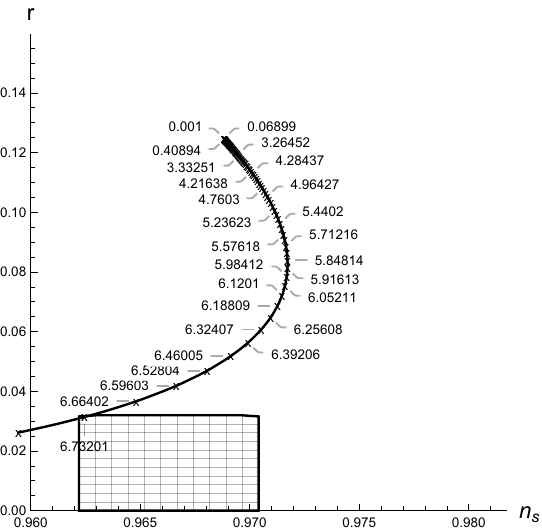}
      \caption{$N=64$}
    \end{subfigure}

    \begin{subfigure}[b]{\textwidth}
      \includegraphics[width=\textwidth]{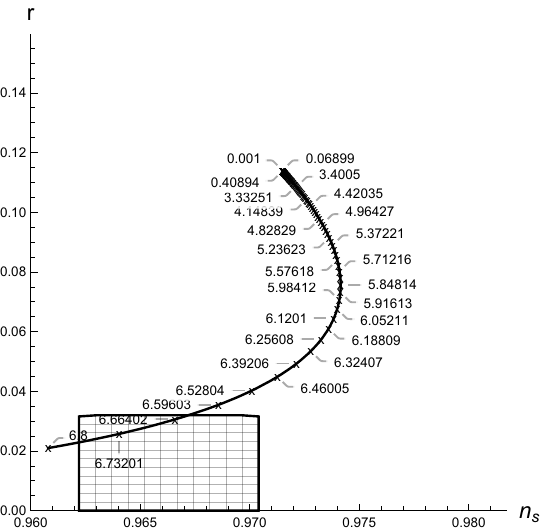}
      \caption{$N=70$}
    \end{subfigure}
  \end{minipage}

  \caption{Potential $V_I$. Inflationary parameters for cases of $N=50$, $60$, $64$, and $70$ e-foldings. Plot markers correspond to different values of mass $m$ evaluated in $\kappa$ units. The shaded region marks the area consistent with the observational data.}\label{fig:VIPlots}
\end{figure}

Finally, we proceed with the analysis for an arbitrary scalar field mass. We numerically study cases of $N=50$, $60$, $64$, and $70$ e-foldings. The results are shown in Figure \ref{fig:VIPlots}. The model is inconsistent with the observational data for the most natural case of $N=60$ e-folding. It develops the correct values of the scalar spectrum tilt $n_s$ for a large band of masses, but the tensor-to-scalar ratio is too big even for large masses. The case of $N=50$ e-foldings shows even more substantial disagreement, as the model develops a suitable scalar spectrum tilt only a narrow band of large masses $m$. This indicates that the model favours the large number of e-foldings. Cases of $N=64$ and $N=70$ e-foldings support that claim. Namely, the $N=70$ case is consistent with the observational data for $ 6.63 \lesssim \tilde m \lesssim 6.76 $. In terms of the Planck mass $m_\text{P} = \sqrt{ \hbar \, c / G_N}$ this corresponds to $ 0.66 \,  m_\text{P} \lesssim m \lesssim 0.67 \, m_\text{P}$. 

We conclude that the model \eqref{V_I} is consistent with the observational data for $N \gtrsim 64$ e-foldings and $ 0.66 \,  m_\text{P} \lesssim m \lesssim 0.67 \, m_\text{P}$. We will study this model in further publications and discuss its implications in Section \ref{Section_Conclusion}.

\subsection{Second model}

For the effective potential \eqref{V_II} read the slow roll parameters:
\begin{align}
  \varepsilon =& \cfrac{ 6 \, \lambda^4 \, \kappa^2  \, \varphi^2 }{\left( 1 + \frac{3}{2} \, \lambda^2 \, \kappa^2  \, \varphi^2 \right)^2 \, \ln^2 \left( 1  + \frac{3}{2} \, \lambda^2 \, \kappa^2\, \varphi^2 \right)} ~ , &  \eta =& \cfrac{8 \, \lambda^2 \, \left( 1 - \frac{3}{2} \, \lambda^2 \, \kappa^2 \, \varphi^2 \right) }{ \left( 1 + \frac{3}{2}\, \lambda^2 \, \kappa^2 \, \varphi^2 \right)^2 \, \ln \left( 1 + \frac{3}{2} \, \lambda^2 \, \kappa^2 \, \varphi^2 \right)} ~ .
\end{align}
The mass parameter $m$ in \eqref{V_II} appears as a multiplier and does not enter the $\ln$ function, so it does not enter the slow roll parameters and does not affect inflation. The value of this mass parameter $m$ is sensitive to the UV structure of the theory since it is obtained by the cut-off regularisation, as discussed above. The fact that $m$ does not enter any observable parameters suggests that this UV sensitivity is negated.

In full analogy with the previous case, it is useful to study limiting cases first. The model has two suitable limits: the decoupling limit $\lambda \to 0$ and the strong coupling limit $\lambda \to \infty$. In the decoupling limit, the model potential remains well-defined and completely reduces to the $\varphi^2$-potential:
\begin{align}
  V_{II} \overset{\lambda\to 0}{\to} \cfrac{m^2\,\varphi^2}{2} \, .
\end{align}
Once again, this case is well-known, and we will not discuss it further. The strong coupling limit exhibits a more interesting behaviour because both potential and slow roll parameters vanish in that limit. In full analogy with the previous case, the model develops a universal behaviour with a weak mass dependence. Moreover, in that limit, all most all initial value of the scalar field is suitable for inflation. In that sense, the model favours large values of $\lambda$ because they make more room for inflation.

Let us proceed with further study of inflationary parameters. The simple form of the effective potential \eqref{V_II} allows one to obtain an analytical solution for the integral present in the number of e-foldings:
\begin{align}
    - \cfrac{\kappa^2}{4}\,\int \cfrac{V(\varphi)}{V'(\varphi)} \, d\varphi = \cfrac{1}{24\, \lambda^2}\,\Bigg[ \frac32\,\lambda^2\,\kappa^2 \varphi^2 + \operatorname{Li}_2\left(-\frac{3}{2} \, \lambda^2\, \kappa^2 \varphi^2\right) - \left( 1 + \frac{3}{2}\,\lambda^2\,\kappa^2\varphi^2\right) \,\ln \left( 1 + \frac{3}{2}\,\lambda^2\,\kappa^2\varphi^2\right) \Bigg] .
\end{align}
Here $\operatorname{Li}_2$ is the dilogarithm. Although it is possible to obtain this analytic formula, it is still impossible to obtain analytic expressions for the scalar field at the beginning and end of inflation due to the presence of transcendental functions.

\begin{figure}
  \centering

  \begin{minipage}[b]{0.49\textwidth}
    \centering
    \begin{subfigure}[b]{\textwidth}
      \includegraphics[width=\textwidth]{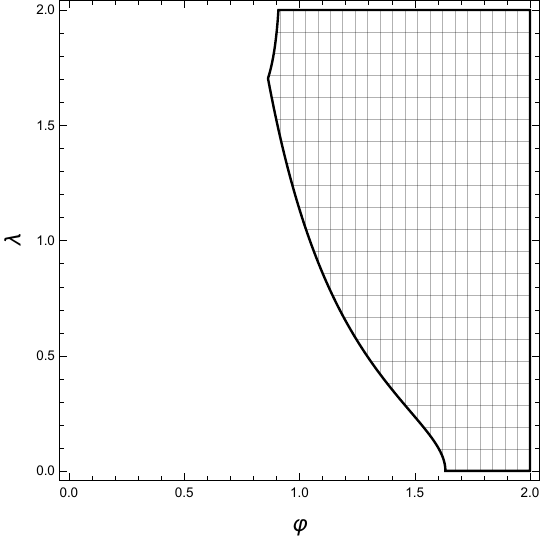}
    \end{subfigure}
  \end{minipage}
  \hfill
  \begin{minipage}[b]{0.49\textwidth}
    \centering
    \begin{subfigure}[b]{\textwidth}
      \includegraphics[width=\textwidth]{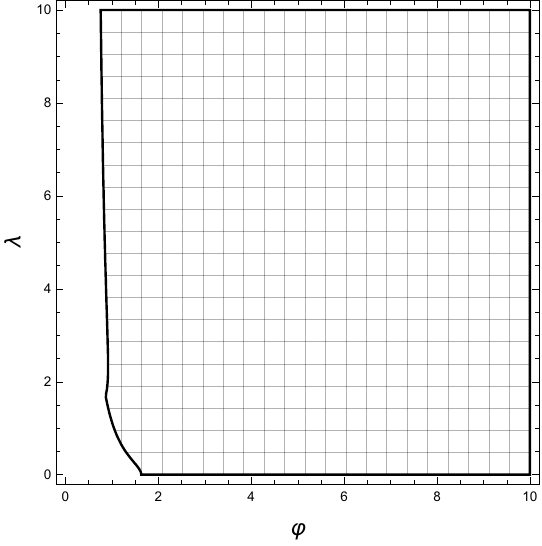}
    \end{subfigure}
  \end{minipage}

  \caption{Potential $V_{II}$. Parameter space of the model. The slow roll parameters are small in the shaded region. Parameter $\varphi$ is given in $\kappa$ units.}\label{fig:VIIParameters}
\end{figure}

\begin{figure}
  \centering
  
  \begin{minipage}[b]{0.49\textwidth}
    \centering
    \begin{subfigure}[b]{\textwidth}
      \includegraphics[width=\textwidth]{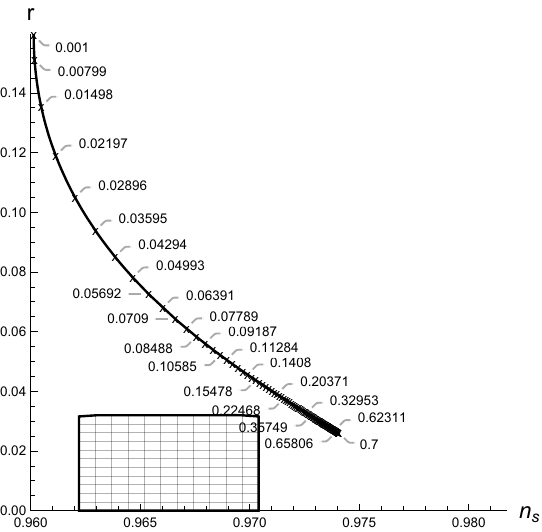}
      \caption{$N=50$}
    \end{subfigure}

    \begin{subfigure}[b]{\textwidth}
      \includegraphics[width=\textwidth]{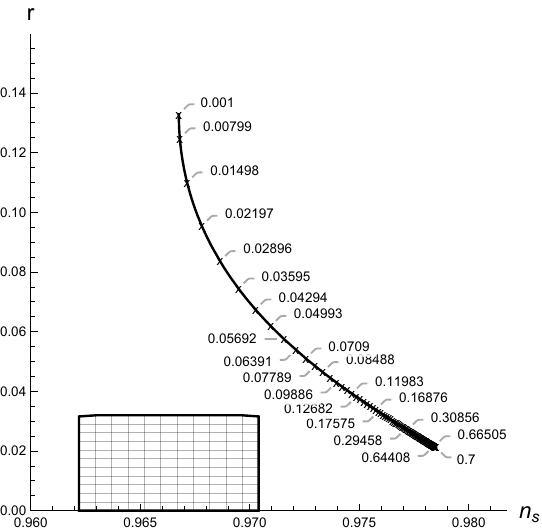}
      \caption{$N=60$}
    \end{subfigure}
  \end{minipage}
  \hfill
  \begin{minipage}[b]{0.49\textwidth}
    \centering
    \begin{subfigure}[b]{\textwidth}
      \includegraphics[width=\textwidth]{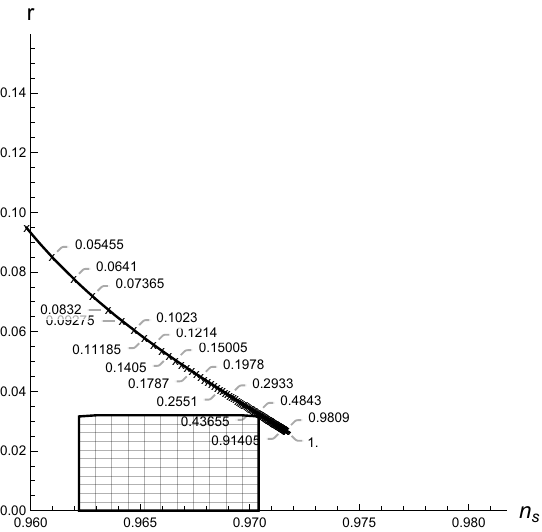}
      \caption{$N=45$}
    \end{subfigure}

    \begin{subfigure}[b]{\textwidth}
      \includegraphics[width=\textwidth]{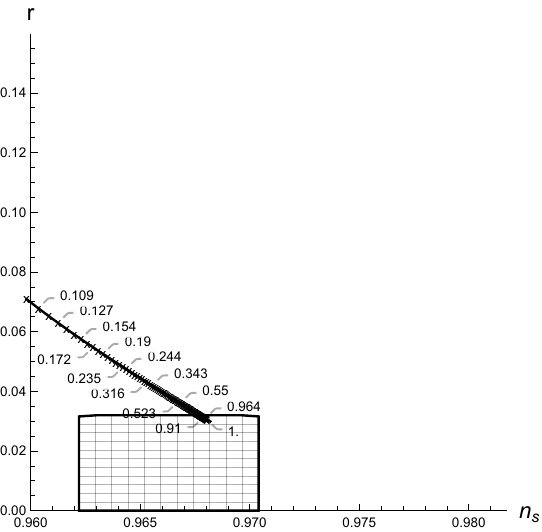}
      \caption{$N=40$}
    \end{subfigure}
  \end{minipage}

  \caption{Potential $V_{II}$. Inflationary parameters for cases of $N=40$, $45$, $50$, and $60$ e-foldings. Plot points correspond to different values of the dimensional parameter $\lambda$. The shaded region marks the area consistent with the observational data.}\label{fig:VIIPlots}
\end{figure}

The area of the model parameter space where the inflationary parameters are small is given in Figure \ref{fig:VIIParameters}. In contrast with the previous case, the region where inflation is allowed has no peculiarities. Inflation parameters for this model for the cases of $N=40$, $50$, and $60$ e-foldings are shown in Figure \ref{fig:VIIPlots}. The model is inconsistent with the observational data for the natural case of $N=60$ e-foldings. Although it admits the correct scalar spectrum tilt for small values of $\lambda$, it develops a large tensor-to-scalar ratio. In contrast with the previous case, the model favours the small number of e-foldings and becomes consistent with the data for $N=40$ e-foldings. At the same time, it also requires a large value of the non-minimal coupling $\lambda \sim 1$.

We conclude that this model is inconsistent with the observational data because it favours a smaller number of e-foldings $N < 50$; it may not last long enough to resolve the horizon and the flatness problems. The model favours the strong coupling regime $\lambda \sim 1$, so it operates at the border of effective field theory applicability. In the strong coupling regime, the original calculations within perturbative quantum gravity can lose their applicability; consequently, the effective potential may no longer be applicable. We discuss the implications of these results further in Section \ref{Section_Conclusion}.

\subsection{Third model}

\begin{figure}
  \centering

  \begin{minipage}[b]{0.4\textwidth}
    \centering
    \begin{subfigure}[b]{\textwidth}
      \includegraphics[width=\textwidth]{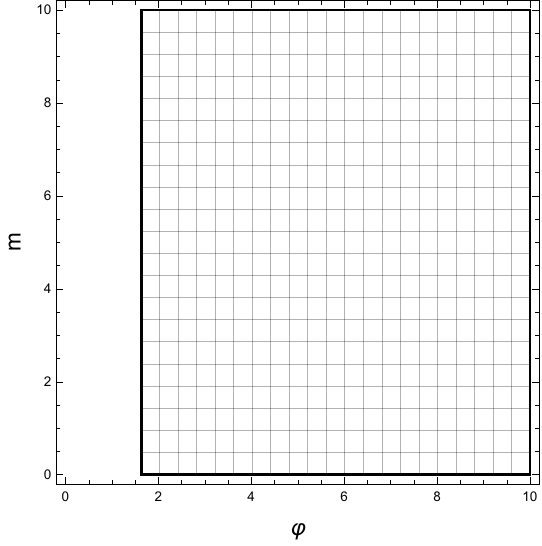}
      \caption{$\lambda=0$, $g=0$.}
    \end{subfigure}

    \vspace{10pt}

    \begin{subfigure}[b]{\textwidth}
      \includegraphics[width=\textwidth]{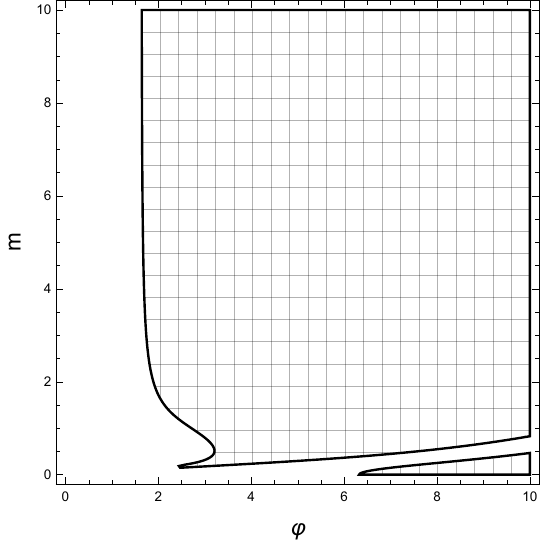}
      \caption{$\lambda=1$, $g=0$.}
    \end{subfigure}

    \vspace{10pt}

    \begin{subfigure}[b]{\textwidth}
      \includegraphics[width=\textwidth]{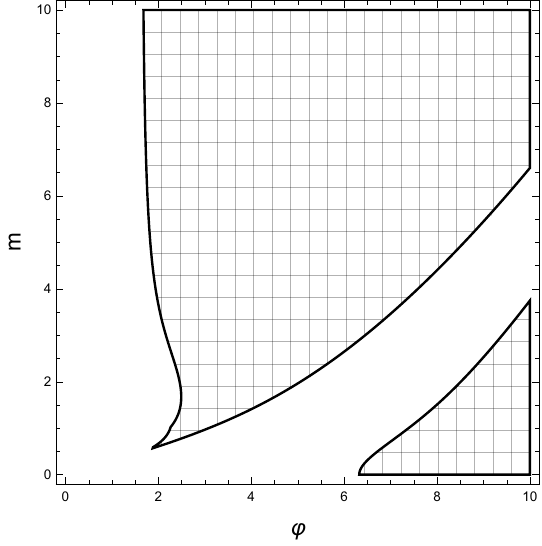}
      \caption{$\lambda=5$, $g=0$.}
    \end{subfigure}
  \end{minipage}
  \hfill
  \begin{minipage}[b]{0.4\textwidth}
    \centering
    \begin{subfigure}[b]{\textwidth}
      \includegraphics[width=\textwidth]{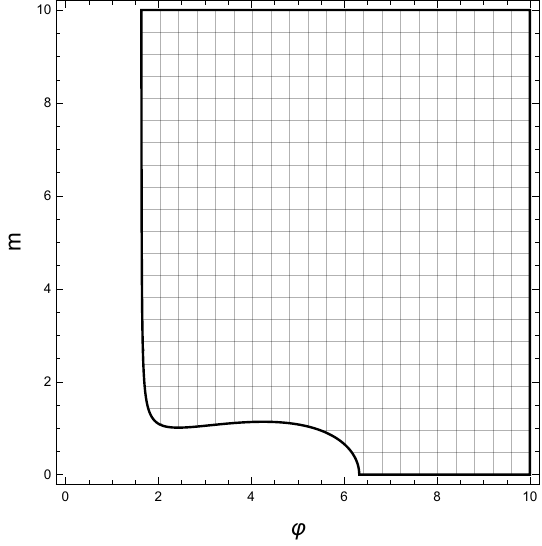}
      \caption{$\lambda=0$, $g=1$.}
    \end{subfigure}

    \vspace{10pt}

    \begin{subfigure}[b]{\textwidth}
      \includegraphics[width=\textwidth]{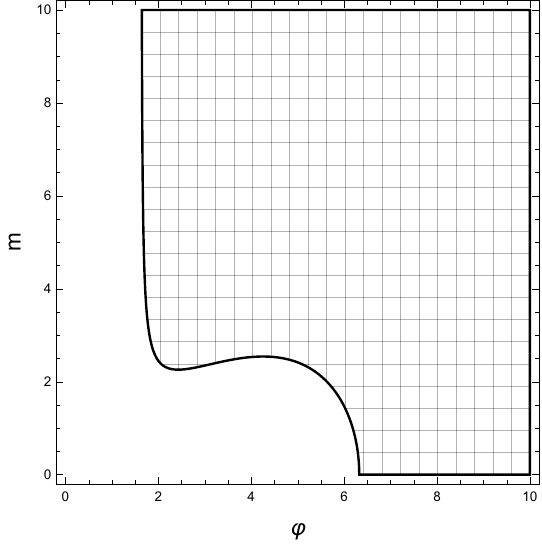}
      \caption{$\lambda=0$, $g=5$.}
    \end{subfigure}

    \vspace{10pt}

    \begin{subfigure}[b]{\textwidth}
      \includegraphics[width=\textwidth]{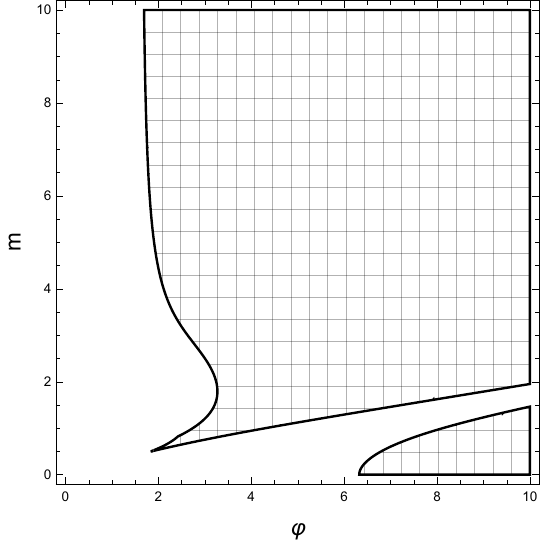}
      \caption{$\lambda=5$, $g=5$.}
    \end{subfigure}
  \end{minipage}

  \caption{Potential $V_{III}$. Slices of parameter space of the model. Dimensional parameters $m$, $\varphi$, and $g$ are given in the $\kappa$ units. The slow roll parameters are small in the shaded region.}\label{fig:VIIIParameters1}
  \label{fig:VIII}
\end{figure}

\begin{figure}
  \centering
  
  \begin{minipage}[b]{0.49\textwidth}
    \centering
    \begin{subfigure}[b]{\textwidth}
      \includegraphics[width=\textwidth]{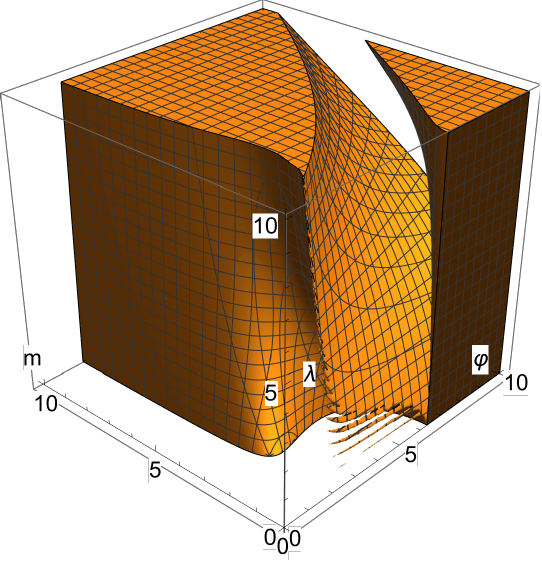}
      \caption{$g=5$.}
    \end{subfigure}
  \end{minipage}
  \hfill
  \begin{minipage}[b]{0.49\textwidth}
    \centering
    \begin{subfigure}[b]{\textwidth}
      \includegraphics[width=\textwidth]{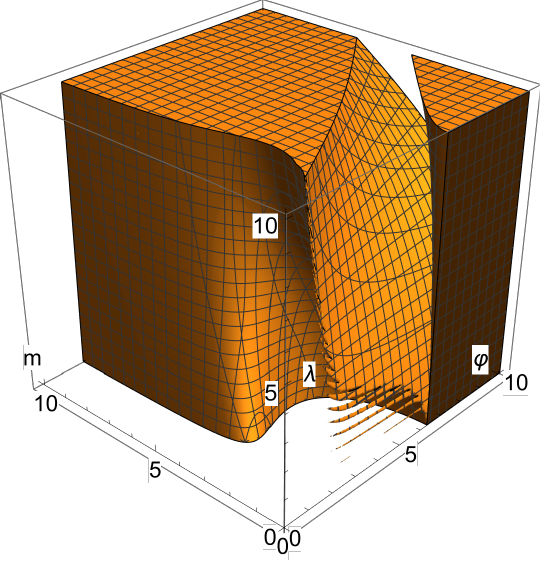}
      \caption{$g=10$.}
    \end{subfigure}
  \end{minipage}

  \caption{Potential $V_{III}$. Parameter space of the model for different values of $g$. Dimensional parameters $m$, $\varphi$, and $g$ are given in the $\kappa$ mass units. The slow roll parameters are small in the shaded region.}\label{fig:VIIIParameters2}
\end{figure}

Finally, we will consider a model with the effective potential \eqref{V_III}. The following expressions give the slow roll parameters:
\begin{align}
    \begin{split}
        \varepsilon =& \cfrac{8}{3\,\kappa^2} \, \cfrac{1}{( 2\,m^2 + \lambda \, \varphi^2 )^2} \, \Bigg[ \Bigg\{ \varphi^7 \, \left( 96 \, \pi^2 \, g \, \lambda \, m^2 + 5 \, \lambda^3 \left( 14 \, \kappa^2 \, m^2 - 9 \,\lambda \right)\right)  + 23040\, \pi^2\, m^6 \\
        & + 6 \, \varphi^5 \, \left( 32 \, \pi^2 \, g \, m^4 + 5\, \lambda^2 m^2 \left( - 9 \, \lambda + 22 \, \kappa^2 \, m^2 + 64\, \pi^2 \right) \right) \varphi \\
        & + 120 \, \lambda \,  m^4 \, \varphi^3 \left( - 3 \, \lambda + 6 \, \kappa^2 \, m^2 + 128\, \pi^2 \right)\\
        & - 60 \,\ln\left( 1 + \cfrac{\lambda\,\varphi^2}{2\,m^2} \right) \, \varphi  \left(2 \, m^3 + \lambda \, m \, \varphi^2 \right)^2 \left( \lambda  \left(\kappa^2 \varphi^2 - 3 \right) + 6 \, \kappa^2 m^2 \right) \Bigg\} \Bigg/ \\
        & \Bigg\{ \varphi^6 \left( 32 \, \pi^2 \, g \, m^2 + 15 \, \lambda^2 \, \left( 2 \, \kappa^2 \, m^2 - \lambda \right)\right) + 180 \left( 64 \, \pi^2 - \lambda \right) m^4 \, \varphi^2  \\
        & + 15 \, \lambda \, m^2 \, \varphi^4 \left( - 9 \, \lambda + 24 \, \kappa^2 \, m^2 + 64 \, \pi^2 \right) - 10 \, \ln\left( 1 + \cfrac{\lambda \,\varphi^2}{2\,m^2} \right) \\
        & \times \left( 36 \, m^6 \, \left( 2 \, \kappa^2 \varphi^2 - 1 \right) + 12 \, \lambda \, m^4 \, \varphi^2 \left( 2 \kappa^2 \, \varphi^2 - 3 \right) + \lambda^2 \, m^2 \, \varphi^4 \, \left( 2 \, \kappa^2 \varphi^2 - 9 \right) \right) \Bigg\} \Bigg]^2, 
    \end{split}
\end{align}
\begin{align}
    \begin{split}
        \eta =& - \cfrac{40}{3\,\kappa^2} \cfrac{1}{(2\,m^2 + \lambda\, \varphi^2)^2} \,\Bigg\{ \varphi^8 \left( 96 \, \pi^2 \, g \, \lambda^2 \, m^2 + \lambda^4 \left( 46 \, \kappa^2 m^2 - 45 \, \lambda \right)\right)\\
        & + 8\, \varphi^6 \left( 48\, \pi^2 \, g \, \lambda \,  m^4 + \lambda^3 \, m^2 \left( - 27 \, \lambda + 44 \, \kappa^2 \, m^2 + 144 \, \pi^2 \right) \right) \\
        & + \varphi^4  \left( 384 \, \pi^2 \, g \, m^6 + 36 \, \lambda^2 \, m^4 \left( - 9 \, \lambda + 22 \, \kappa^2 \, m^2 + 192 \, \pi^2 \right) \right) + 9216 \, \pi^2 \, m^8 \\
        & + 144 \, \lambda \,  m^6 \, \varphi^2 \, \left( - \lambda + 2 \, \kappa^2 \, m^2 + 96 \, \pi^2 \right) - 12 \,\ln\left( 1 + \cfrac{\lambda\,\varphi^2}{2\, m^2} \right) \\
        & \times \left( 2 \, m^3 + \lambda \, m \, \varphi^2 \right)^2 \left( \lambda^2 \, \varphi^2  \left( 5 \, \kappa^2 \, \varphi^2 - 9 \right) + 12 \, \kappa^2 \, m^4 + 6\, \lambda \, m^2 \left( 4 \, \kappa^2  \varphi^2 - 1 \right)\right) \Bigg\} \Bigg/\\
        & \Bigg\{ \varphi^6 \left( 15 \, \lambda^2 \left( \lambda - 2 \, \kappa^2 \, m^2 \right) - 32 \, \pi^2 \, g \, m^2 \right) + 180 \, \left( \lambda - 64\, \pi^2 \right) \, m^4 \, \varphi^2 \\
        & - 15 \, \lambda \, m^2 \, \varphi^4 \left( - 9 \, \lambda + 24 \, \kappa^2 \, m^2 + 64 \, \pi^2 \right) + 10 \, \ln\left( 1 + \cfrac{\lambda \, \varphi^2}{2\,m^2} \right) \\
        & \times \left( 36 \, m^6 \left( 2 \, \kappa^2 \varphi^2 - 1 \right) + 12 \, \lambda \, m^4 \, \varphi^2 \left( 2 \, \kappa^2 \varphi^2 - 3 \right) + \lambda^2 \, m^2 \, \varphi^4  \left( 2 \, \kappa^2 \varphi^2 - 9 \right) \right) \Bigg\} .
    \end{split}
\end{align}
\noindent Figures \ref{fig:VIIIParameters1} and \ref{fig:VIIIParameters2} contain plots showing regions of the model parameter space where these slow roll parameters are small.

\begin{figure}
  \centering

  \begin{minipage}[b]{0.45\textwidth}
    \centering
    \begin{subfigure}[b]{\textwidth}
      \includegraphics[width=\textwidth]{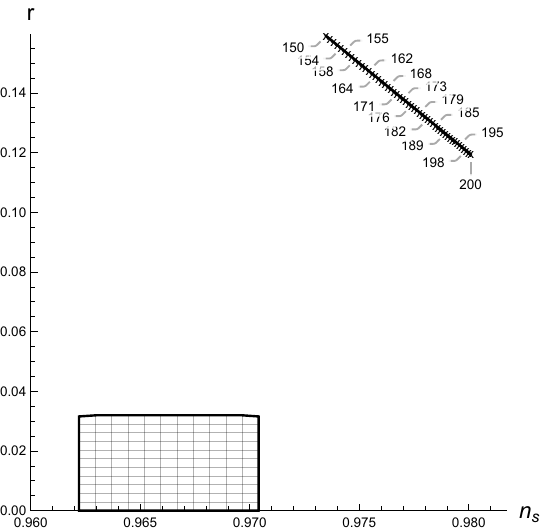}
      \caption{Massless limit}
    \end{subfigure}
  \end{minipage}
  \hfill
  \begin{minipage}[b]{0.45\textwidth}
    \centering
    \begin{subfigure}[b]{\textwidth}
      \includegraphics[width=\textwidth]{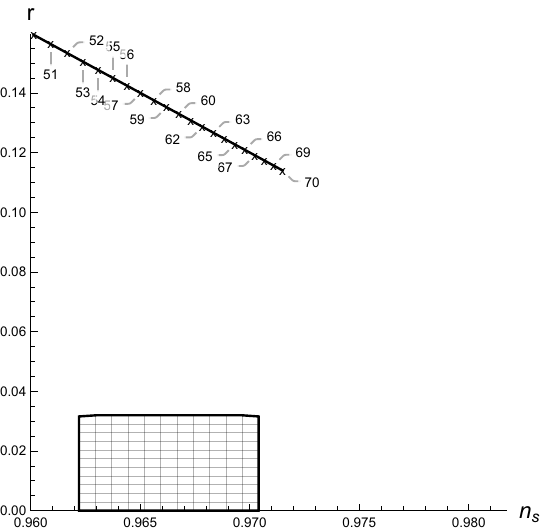}
      \caption{Decoupling limit}
    \end{subfigure}
  \end{minipage}

  \caption{Potential $V_{III}$. Inflationary parameters for the limiting cases. Plot markers correspond to different numbers of e-folding. The shaded region marks parameters consistent with the observational data.}\label{fig:VIIIlimiting}
\end{figure}

The model admits two important liming cases: the massless case $m\to 0$ and the decoupling case $\lambda \to 0$, $ g \to 0$. In the massless case, the slow roll parameters read
\begin{align}
  \varepsilon & = \cfrac{24}{\tilde\varphi^2}\, , & \eta & = \cfrac{40}{\tilde\varphi^2} \,.
\end{align}
Here we use the same definition of the dimensionless scalar field $\tilde\varphi$. This gives the scalar field value at the end of inflation
\begin{align}
  \tilde\varphi_\text{end} = 2\, \sqrt{10}.
\end{align}
The initial value of the scalar field is related to the number of e-foldings $N$:
\begin{align}
  \varphi_\text{init} = \sqrt{48\, N + 40} \,.
\end{align}
Which gives the following relations for the inflationary parameters:
\begin{align}
  r & = \cfrac{144}{6\,N + 5 } \, , & n_s & = \cfrac{6\,N -19 }{6\, N + 5} \, .
\end{align}
In this limit, the model is inconsistent with the observational data. Namely, the tensor-to-scalar ratio is small enough for $ N \gtrsim 750$ while the scalar spectrum tilt is consistent for $110 \lesssim N \lesssim 134$. In the decoupling limit $\lambda \to 0$, $g \to 0$, the model potential is reduced to $\varphi^2$ potential if full similarity with previous cases. Plots showing the model behaviour in these limiting cases are given in Figure \ref{fig:VIIIlimiting}.

These limiting cases have a certain similarity with the first model. For both models, the limiting cases are inconsistent with the data. Nonetheless, the first model is consistent with the data for some intermediate parameter values. Therefore, the discussed model \eqref{V_III} can only be consistent with the data for some intermediate parameter values. The search for such a parameter region is a challenging problem because of the structure of parameter space. Figures \ref{fig:VIIIParameters1} and \ref{fig:VIIIParameters2} show that the acceptable region of parameter space is split into two pieces. Still, it is possible to constrain the problem reasonably and to perform some meaningful analysis.

We base our treatment on the following premises. Firstly, we are only interested in a reasonably small range of masses $ \kappa \, m <10$. These masses lie below the Planck mass $m_\text{P}$, so they do present immediate interest. However, it may be possible to justify the usage of models with the mass of a scalar field larger than the Planck mass. We believe this case is irrelevant within the effective field theory treatment where the discussed effective potential is obtained. Secondly, we assume that the initial value of the scalar field should be large. Following the well-known description of inflation \cite{Linde:1983gd}, it is safe to assume that the initial values of the scalar field lie in the Planck region, as at the beginning of inflation, the universe is created from a high energy quantum state. These two considerations suggest that the inflation shall occur in the region which in Figure \ref{fig:VIIIParameters1} is situated in the bottom right corner.

The structure of that region favours two regimes: $\lambda =0 $ and $\lambda \gg 1$, $\tilde{g} \sim 0$ where $\tilde{g}$ is the dimensional coupling $g$ in $\kappa$ units. In the $\lambda=0$ case, the region is not split into two parts for any $\tilde{g}$ values and can be analysed easily. The region is split into two parts if $\lambda \sim 0$ and $g\sim 0$. However, the desired region enforces extremely large initial and final scalar field values, so it can hardly be relevant. The situation is similar for $\lambda \gg 1$ and $\tilde{g} \gg 1$. The only remaining case, $\lambda \gg 1$, $\tilde{g} \sim 0$, admits scenarios with reasonably large initial and final values of the inflationary parameters.

\begin{figure}
  \centering
  
  \begin{minipage}[b]{0.45\textwidth}
    \centering
    \begin{subfigure}[b]{\textwidth}
      \includegraphics[width=\textwidth]{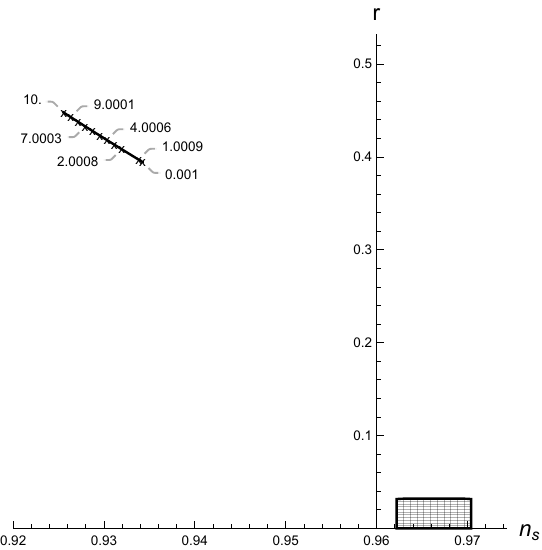}
      \caption{$g=1$}
    \end{subfigure}
  \end{minipage}
  \hfill
  \begin{minipage}[b]{0.45\textwidth}
    \centering
    \begin{subfigure}[b]{\textwidth}
      \includegraphics[width=\textwidth]{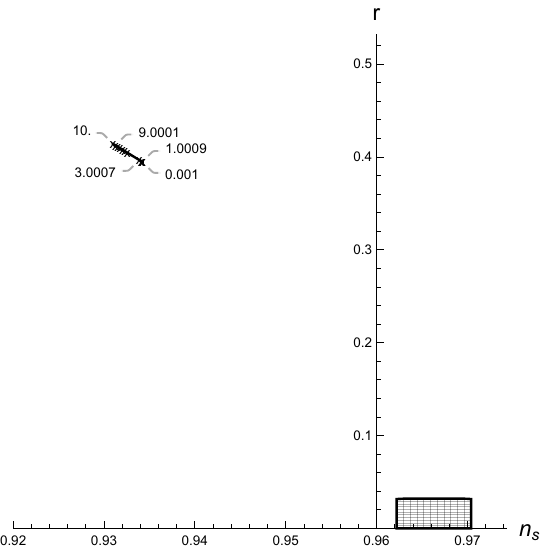}
      \caption{$g=10$}
    \end{subfigure}
  \end{minipage}
  
  \caption{Potential $V_{III}$ with $\lambda=0$. Plot markers correspond to different values of the mass parameter $m$. Parameters $\varphi$, $g$ and $m$ are given in $\kappa$ mass units.} \label{fig:VIIIL0}
\end{figure}

\begin{figure}
  \centering

  \begin{minipage}[b]{0.45\textwidth}
    \centering
    \begin{subfigure}[b]{\textwidth}
      \includegraphics[width=\textwidth]{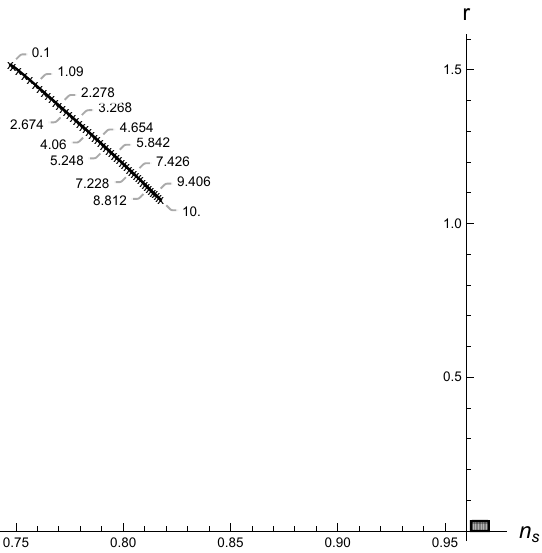}
      \caption{$\lambda=1$}
    \end{subfigure}
  \end{minipage}
  \hfill
  \begin{minipage}[b]{0.45\textwidth}
    \centering
    \begin{subfigure}[b]{\textwidth}
      \includegraphics[width=\textwidth]{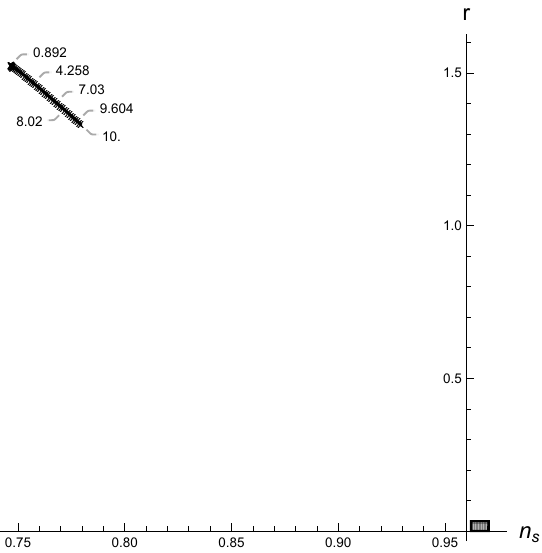}
      \caption{$\lambda=5$}
    \end{subfigure}
  \end{minipage}

  \caption{Potential $V_{III}$. Inflationary parameters for the limiting cases $g = 0 $ for different values of $\lambda$. Plot markers correspond to different numbers of mass $m$. The shaded region marks parameters consistent with the observational data. Dimensional parameters $m$, $\varphi$, and $g$ are given in the $\kappa$ mass units.} \label{fig:VIIIg0}
\end{figure}

Let us start with the $\lambda =0 $ case. It is inconsistent with the observational data. Figure \ref{fig:VIIIL0} presents plots for $g=1$ and $g=10$ cases. These plots show that the inflationary parameters approach the desired region for smaller values of masses. At the same time, as discussed above, the massless limit is also inconsistent with the data by a significant margin. Thus, we conclude that this case can hardly be brought in consistence with the data. The other case, $\lambda \gg 1$, $\tilde{g} =0$, is also inconsistent with the data. Figure \ref{fig:VIIIg0} shows inflationary parameters for different masses with $\lambda = 1$ and $\lambda = 5$.

Based on this data, there is no reason to believe that the model admits a set of parameters consistent with the observational data. Let us highlight one more time that the chosen region of parameters corresponds to the most natural physical setup. Although we cannot exclude the possibility of tuning this model into an agreement with the data, such a set of parameters would be highly marginal. We discuss these results in Section \ref{Section_Conclusion}.

\section{Discussion and conclusion}\label{Section_Conclusion}

We explored the possibility of describing inflation as an effect dynamically generated by quantum effects at the one-loop level. We studied three simplest models proposed in papers \cite{Arbuzov:2020pgp,Arbuzov:2021yai}. 

The first model describes a single scalar field of non-vanishing mass minimally coupled to general relativity. The model develops an effective potential \eqref{V_I}. Plots of its inflation parameters for $N=50$, $60$, $64$, and $70$ e-foldings are shown in Figures \ref{fig:VIPlots}. The model is consistent with the observational data for $N \gtrsim 64$ e-foldings. The corresponding suitable values of the mass lie beyond the Planck scale. Namely, for $N=70$ the allowed mass rage is $ 0.66 \, m_\text{P} \lesssim m < 0.67 \, m_\text{P} $.

The second model describes a single massless scalar field non-minimally coupled to the Einstein tensor. The model develops an effective potential \eqref{V_II}, generating the scalar field mass. The generated mass parameter is UV sensitive because the effective potential is evaluated in the cut-off regularisation scheme. In other words, the new mass parameter depends on the cut-off scale of the model. This UV dependence is completely negated because the mass parameter does not enter any expression for the inflation parameters and does not affect inflation. Plots of the inflation parameters of the model for $N=40$, $45$, $50$, and $60$ e-foldings are shown in Figure \ref{fig:VIIPlots}. The model is disfavoured because it is consistent with the data only for a small number of e-foldings and in the strong coupling regime. Because of the small number of e-foldings, the model may not be capable of resolving the horizon and the flatness problems. Because the model requires a strong coupling regime, the expression for the effective potential calculated within perturbative quantum gravity may no longer be applicable.

The last model describes a generalisation of the Coleman-Weinberg model for the gravitational case. The model develops an effective potential \eqref{V_III}. It has three parameters, and only one of them, the six-particle coupling $g$, is UV sensitive. The model is inconsistent with the observational data for the realistic range of parameter values. Plots of the inflation parameters of this model are shown in Figures \ref{fig:VIII}.

We interpret these results in the following way. As discussed in the introduction, these models present explicit examples of inflationary scenarios driven purely by quantum effects. They are similar to Starobinsky inflation, which uses higher curvature terms to describe inflationary expansion. The models discussed in this paper use a similar approach to inflation, but use the scalar field effective potential to drive inflation. Each model provides a separate piece of information about the viability of such an approach to inflation.

The first model provides the best example of inflation driven by pure quantum effects. The model relies on the simplest possible model of quantum scalar-tensor gravity, yet it can generate an effective potential consistent with the observations. The model has a single free parameter which is the scalar field mass. In turn, the mass is well-constrained for the model to be consistent and lies below the Planck scale.

The second model shows the role of possible non-minimal coupling to gravity. The model can be consistent with the data but for marginal values of parameters. The model favours a small number of e-foldings and a strong coupling regime. It is safe to conjecture that if such a non-minimal coupling is introduced at the level of microscopic action together with the mass term, the resulting model will improve the first model.

The third model shows that the scalar field potential presented in the microscopic action plays the leading role. The Coleman-Weinberg model on its own is inconsistent with the Planck data \cite{Barenboim:2013wra}. Its generalisation considered in this paper admits a larger space of parameters, but it is still inconsistent with the observations for the natural choice of parameters. This case shows that quantum gravitational corrections can hardly significantly improve a potential initially inconsistent with the data.

Lastly, the following comment is due to terms omitted from the effective action. The full effective action includes the effective potential and non-potential terms. The most well-known is the non-minimal kinetic coupling between a scalar field and the Einstein tensor. It is universally generated at the one-loop level \cite{Latosh:2020jyq} and can also drive inflation \cite{Kobayashi:2011nu,Kobayashi:2010cm}. Therefore, it is essential to develop the discussed approach to inflation and consider models with non-minimal couplings. This development will be done in further publications.

\section*{Acknowledgement}

The work by AA and DK was supported by RSF grant 22-22-00294. BL work was supported by the Institute for Basic Science Grant IBS-R018-Y1.

\bibliographystyle{unsrturl}
\bibliography{main_arbuzov_2023.bib}

\newpage

\begin{figure}
  \centering

  \begin{minipage}[b]{0.49\textwidth}
    \centering
    \begin{subfigure}[b]{\textwidth}
      \includegraphics[width=\textwidth]{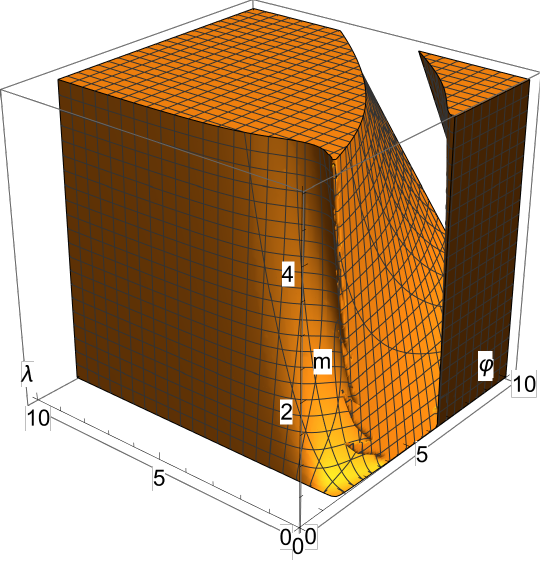}
      \caption{$g=0$.}
    \end{subfigure}
  \end{minipage}
  \hfill
  \begin{minipage}[b]{0.49\textwidth}
    \centering
    \begin{subfigure}[b]{\textwidth}
      \includegraphics[width=\textwidth]{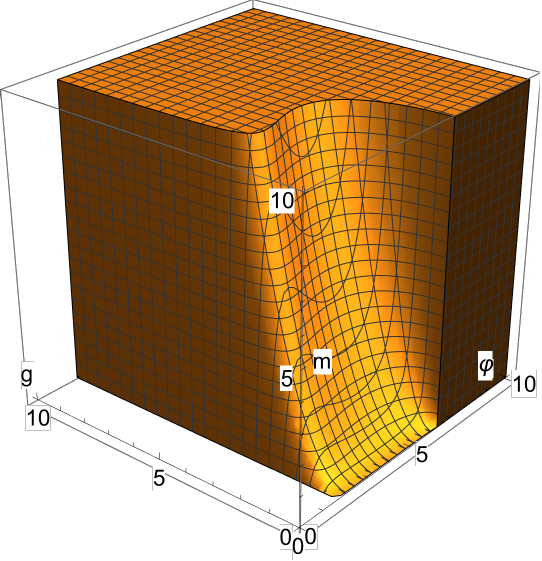}
      \caption{$\lambda=0$.}
    \end{subfigure}
  \end{minipage}
  \caption{Potential $V_{III}$. Parameter space of the model for limiting cases. Dimensional parameters $m$, $\varphi$, and $g$ are given in the $\kappa$ mass units. The slow roll parameters are small in the shaded region.}
\end{figure}

\end{document}